\documentclass[aps,12pt,final,notitlepage,oneside,onecolumn,nobibnotes,nofootinbib,superscriptaddress,noshowpacs,centertags]{revtex4}
\usepackage{graphicx}
\usepackage{graphics}
\usepackage{epsfig}
\usepackage{verbatim}
\usepackage{amssymb}
\usepackage{amsmath}
\usepackage{amsfonts}

\begin{document}
\selectlanguage{english}

\title{Microscopic Description of Isoscalar Giant Monopole Resonance in $^{48}$Ca}

\author{\firstname{N.~N.}~\surname{Arsenyev}}
\email{arsenev@theor.jinr.ru}
\affiliation{%
Bogoliubov Laboratory of Theoretical Physics,
             Joint Institute for Nuclear Research,
             141980 Dubna, Moscow region, Russia
}%
\author{\firstname{A.~P.}~\surname{Severyukhin}}
\email{sever@theor.jinr.ru}
\affiliation{%
Bogoliubov Laboratory of Theoretical Physics,
             Joint Institute for Nuclear Research,
             141980 Dubna, Moscow region, Russia
}%
\affiliation{%
Dubna State University, 141982 Dubna, Moscow region, Russia
}%

\begin{abstract}
{\bf Abstract -} The properties of the isoscalar giant monopole resonance (ISGMR)
for the double magic $^{48}$Ca are analyzed in the framework of
a microscopic model based on Skyrme-type interactions.
A method for simultaneously taking into account the coupling
between one-, two-, and three-phonon terms
in the wave functions of $0^{+}$~states has been developed.
The inclusion of three-phonon configurations leads to
a substantial redistribution of the ISGMR strength to
lower energy $0^{+}$~states and also higher energy tail.
Our results demonstrate that the developed approach enables
to us to describe a gross structure of the ISGMR spreading width.
\end{abstract}

\maketitle

%
%
\section{Introduction}

The atomic nucleus is a most exceptional quantum many-body system
which is characterized by the appearance of various collective excitations.
As one of the fundamental excitations in a nucleus, giant
resonances are small-amplitude collective vibration modes~\cite{Bortignon1998,Harakeh2001}.
Among the giant resonances, a special place is
occupied by the isoscalar giant monopole resonances (ISGMR).
The ISGMR, the so-called breathing mode, is well localized,
which makes it possible to judge from the excitation energies
about the incompressibility coefficient of nuclear
matter $K_{\infty}$~\cite{Bohigas1979,Blaizot1980}. The incompressibility or
the compression modulus is a very fundamental quantity
closely connected to the saturation properties of nuclei and to
the nucleon--nucleon interaction~\cite{Blaizot1980}. This parameter of
the nuclear equation of state is important in several physical
contexts such as prompt supernova explosions~\cite{Baron1985,Bethe1990},
the interiors of neutron stars~\cite{Glendenning1986,Lattimer2001},
heavy-ion collisions at intermediate and high
energies~\cite{Stocker1986,Baran2005}.

The extrapolation to nuclear matter and neutron stars requires that
the energy of the ISGMR has to be known over a wide range of
the mass number~$A$. In addition to the above, it is also well known
that the finite nucleus incompressibility $K_{A}$~exhibits
a strong mass dependence~\cite{Blaizot1980,Blaizot1976}.
Thus, experimental information
on the ISGMR from a wide range of nuclei is important.
The ISGMR is a well-defined experimental observable, which can be
measured precisely through various experimental techniques.
The most successful experimental approach~\cite{Harakeh2001} to study
the ISGMR is inelastic scattering at extreme forward angles because
of its selectivity to isoscalar excitations and the dominance
of the monopole cross section at angles close to $0^{\circ}$.
The ISGMR has been established in heavy nuclei ($A>90$) where it forms
a compact resonance with a Lorentzian shape (see~\cite{Garg2018}, and
references therein). It is a general problem in $A<90$ nuclei that
the isoscalar $E0$~strength was found to be very fragmented and no longer
concentrated in one single peak. Recently, results have become available
from the Research Center for Nuclear Physics (RCNP) \cite{Howard2020} and
the iThemba Laboratory for Accelerator
Based Sciences (iThemba LABS) \cite{Olorunfunmi2022} groups for $^{40,42,44,48}$Ca
obtained through small angle (including $0^{\circ}$) inelastic $\alpha$-scattering
measurements at 386 and 196~MeV, respectively. Both experimental datasets show
the fragmentation and splitting of the ISGMR strength distributions. While the
iThemba LABS results only cover the range 9.5--25.5~MeV, the studies
performed at RCNP presented $E0$~strengths for significantly
larger excitation-energy range 10--31~MeV. For the three nuclei
$^{40,42,44}$Ca one observes that the monopole strength distributions
from the iThemba LABS study are weaker than those from RCNP, especially
at excitation energies around and below the peak of the distribution;
see Fig.~6 in~\cite{Olorunfunmi2022}. For the case of $^{48}$Ca,
in an early $(\alpha,\alpha^{\prime})$ study performed by
the Texas A\&M University Cyclotron Institute (TAMU) group~\cite{Lui2011}
presented the isoscalar $E0$ strength distribution, which was located between
9.5 and 40~MeV. The iThemba LABS results are in better agreement
with previous RCNP datasets, although still weaker than the results
from TAMU below the peak of the distribution; see Fig.~6
in~\cite{Olorunfunmi2022}. In general, the experimental findings
for the $^{40,42,44,48}$Ca isotopes are close. It should be noted that
the ISGMR strength distributions were measured using
the high energy-resolution capabilities at RCNP, iThemba LABS
and TAMU. This allows for the observation of pronounced
fine structure. A viable way of understanding the origin and
nature of fine structure is to extract the characteristic energy
scales from the experimental data and compare them with
the most suitable theoretical models.

On the theoretical side, the standard tools for the fine-structure
analysis of the ISGMR are random phase approximation~(RPA)
and its quasiparticle version~(QRPA) in the case of open-shell nuclei.
It means that the collective $0^{+}$~states are constructed
in mean-field theory as a coherent superposition of one-particle--one-hole
($1p1h$) excitations~\cite{Rowe1970,Ring1980}. The approach employs
the self-consistent mean-field derived from effective nucleon--nucleon
interactions that are taken as nonrelativistic two-body Gogny
\cite{Blaizot1976,Blaizot1977,Peru2005,Gambacurta2012} and Skyrme
\cite{Adachi1978,Giai1981a,Agrawal2004,Terasaki2005,Tselyaev2009,Khan2013,RocaMaza2017,Severyukhin2017,Gambacurta2019,Colo2020,Arsenyev2021}
forces. They are also derived from relativistic Lagrangians, e.g.,
\cite{Vretenar2003,Litvinova2007,Piekarewicz2015,Piekarewicz2017}.
For spherical nuclei, the extension of the wave function to
more complex configurations increases the fragmentation of
the initial $1p1h$~doorway states over many excited states
and determines the damping of the giant resonances.
As pointed out in
\cite{Gambacurta2012,Adachi1978,Tselyaev2009,RocaMaza2017,Severyukhin2017,Gambacurta2019,Arsenyev2021,Litvinova2007,Drozdz1990,Kamerdzhiev2004},
it allows us to discuss the fine structure of the ISGMR and its damping properties.
In Refs.~\cite{Severyukhin2017,Arsenyev2021}
we have shown that the calculations taking into account the phonon--phonon coupling~(PPC)
describes reasonably well the gross structure of spreading widths of
giant monopole resonances in the heavy and superheavy nuclei.
In~\cite{Olorunfunmi2020b} it was shown that the PPC predictions
of the fine structure of the ISGMR in $^{48}$Ca are in good agreement
with the fine structure which is extracted from experimental data analysis.
The PPC calculations use a formalism in which the RPA phonons
are treated as boson excitations~\cite{Soloviev1992}.
The two-phonon amplitudes are defined by the coupling between $1p1h$
and the two-particle--two-hole ($2p2h$) excitations.
These are the basic ingredients of the quasiparticle--phonon model
(QPM), but the single-particle~(SP) spectrum and
the residual interaction are calculated with the Skyrme force;
see, for example,~\cite{Severyukhin2004,Severyukhin2012}.

It is remarkable that thanks to experimental advancements,
especially at rare ion beam facilities, the investigation of
ISGMR of proton- and neutron-rich unstable
nuclei has become possible. For exotic nuclei
with a large neutron excess, a soft monopole mode
may emerge, which brings new insights into the nuclear incompressibility and
has become the goals for both experimental and theoretical investigations.
For instance, it has been observed experimentally in $^{11}$Li~\cite{Fayans1992}
and $^{68}$Ni~\cite{Vandebrouck2014}, and is predicted in
the neutron-rich Ca~\cite{Piekarewicz2017} and
Ni~\cite{Piekarewicz2015}, Sn and Pb~\cite{Khan2013} isotopes.
However, for heavy and deformed exotic nuclei, the structure
and mechanism of the soft monopole mode are not clear.
The soft monopole strength becomes much more fragmented
when beyond-mean-field effects are considered~\cite{Gambacurta2019}.
However, most of the available theoretical calculations are
based on the QRPA or beyond QRPA which is limited to interaction
with the $2p2h$~configurations.

In the present paper, we propose an extension of our PPC approach
taking into account the coupling between one-, two-, and three-phonon
terms in the wave functions of excited $0^{+}$~states.
As an application of the method we present results for
the change in the monopole strength function profile caused by
the $3p3h$~fragmentation and analyze its impact on
the characteristics of the ISGMR in the case of double closed-shell
nucleus $^{48}$Ca.
The recent ISGMR strength distribution measurements through small
angle inelastic $\alpha$-scattering give an opportunity to compare our results
and the experimental data~\cite{Howard2020,Olorunfunmi2022}.

%
%
\section{Basic elements of the microscopic approach}

The starting point of our microscopic approach consists
in the Hartree--Fock~(HF) calculation of the ground state
based on the Skyrme energy density functional~(EDF) \cite{Ring1980,Bender2003}.
Spherical symmetry is imposed on the HF wave functions.
The continuous part of the SP spectrum is discretized by
diagonalizing the HF Hamiltonian on a harmonic
oscillator basis~\cite{Blaizot1977}. The residual
particle--hole interaction is obtained as the second
derivative of the energy density functional with respect to
the particle density. By means of the standard
procedure~\cite{Terasaki2005} we obtain the familiar RPA equations in
the $1p1h$~configuration space. The eigenvalues of the RPA
equations are found numerically as the roots of a relatively
simple secular equation within the finite rank separable
approximation~(FRSA)~\cite{Giai1998,Severyukhin2008}.
Making use of the FRSA for the residual interaction enables
us to perform RPA calculations in very large configurational
spaces. In particular, the cutoff of the discretized continuous
part of the SP spectra is at the energy of 100~MeV. This is
sufficient to exhaust practically all the energy-weighted
sum rules~(EWSR) within the RPA. It is worth mentioning that
the so-called FRSA has been successfully used
to study the electromagnetic transitions between excited
states~\cite{Severyukhin2012,Severyukhin2020}
and the ISGMR strength distribution within and beyond the
QRPA~\cite{Severyukhin2017,Arsenyev2021,Olorunfunmi2020b}.

Using the basic QPM ideas in the simplest case of the
configuration mixing~\cite{Soloviev1992,LoIudice2012},
we construct the wave functions from a linear combination
of one-, two-, and three-RPA phonons \cite{Grinberg1994,Grinberg1998,Severyukhin2021},
\begin{equation}
  \begin{split}
  \Psi_{\nu}(JM)=\Biggl(&\sum\limits_{i}R_{i}(J\nu)Q_{JMi}^{+}
  +\sum\limits_{\lambda_{1}i_{1}\lambda_{2}i_{2}}
  P_{\lambda_{2}i_{2}}^{\lambda_{1}i_{1}}(J\nu)
  \left[Q_{\lambda_{1}\mu_{1}i_{1}}^{+}Q_{\lambda_{2}\mu_{2}i_{2}}^{+}\right]_{JM}\\
  &+\sum\limits_{\lambda_{1}i_{1}\lambda_{2}i_{2}\lambda_{3}i_{3}{}J^{\prime}}
  T_{J^{\prime}{\,}\lambda_{3}i_{3}}^{\lambda_{1}i_{1}{\,}\lambda_{2}i_{2}}(J\nu)
  \left[\left[Q_{\lambda_{1}\mu_{1}i_{1}}^{+}Q_{\lambda_{2}\mu_{2}i_{2}}^{+}\right]_{J^{\prime}}
  Q_{\lambda_{3}\mu_{3}i_{3}}^{+}\right]_{JM}\Biggr)|0\rangle,
  \label{wf3ph}
  \end{split}
\end{equation}
in which $\lambda$ denotes the total angular momentum,
$\mu$ is its $z$-projection in the laboratory system and
sequential number $i$; the $[\ldots]_{JM}$ stands for angular momentum
coupling. The ground state is the RPA phonon
vacuum $|0\rangle$. The wave functions of the one-RPA phonon
excited states, having energy $\omega_{\lambda i}$, given
by $Q_{\lambda\mu i}^{+}|0\rangle$ as a superposition of the
$1p1h$~configurations. The normalization condition for
the wave functions~(\ref{wf3ph}) yields
\begin{equation}
  \sum\limits_{i}\left[R_{i}(J\nu)\right]^{2}+2\sum\limits_{\lambda_{1}i_{1}\\\lambda_{2}i_{2}}
  \left[P_{\lambda_{2}i_{2}}^{\lambda_{1}i_{1}}(J\nu)\right]^{2}
  +6\sum\limits_{\lambda_{1}i_{1}\lambda_{2}i_{2}\lambda_{3}i_{3}{\,}J^{\prime}}
  \left[T_{J^{\prime}{\,}\lambda_{3}i_{3}}^{\lambda_{1}i_{1}{\,}\lambda_{2}i_{2}}(J\nu)\right]^{2}=1.
  \label{wf3phnorm}
\end{equation}
The variational principle leads to a set of linear equations for
the unknown amplitudes $R_{i}(J\nu)$, $P_{\lambda_{2}i_{2}}^{\lambda_{1}i_{1}}(J\nu)$ and
$T_{J^{\prime}{\,}\lambda_{3}i_{3}}^{\lambda_{1}i_{1}{\,}\lambda_{2}i_{2}}(J\nu)$ (see details
in~\cite{Soloviev1992,LoIudice2012,Grinberg1994,Grinberg1998}):
\begin{equation}
  R_{i}(J\nu)\left(\omega_{Ji}-E_{\nu}\right)
  +\sum\limits_{\lambda_{1}i_{1}\lambda_{2}i_{2}}
  P_{\lambda_{2}i_{2}}^{\lambda_{1}i_{1}}(J\nu)
  U_{\lambda_{2}i_{2}}^{\lambda_{1}i_{1}}(Ji)=0,
  \label{3pheq1}
\end{equation}
\begin{equation}
  \begin{split}
  P_{\lambda_{2}i_{2}}^{\lambda_{1}i_{1}}(J\nu)
  \left(\omega_{\lambda_{1}i_{1}}+\omega_{\lambda_{2}i_{2}}-E_{\nu}\right)
  &+3\sum\limits_{\lambda_{1}^{\prime}i_{1}^{\prime}\lambda_{2}^{\prime}i_{2}^{\prime}}
  T_{\lambda_{1}{\,}\lambda_{2}i_{2}}^{\lambda_{1}^{\prime}i_{1}^{\prime}{\,}
  \lambda_{2}^{\prime}i_{2}^{\prime}}(J\nu)
  U_{\lambda_{2}^{\prime}i_{2}^{\prime}}^{\lambda_{1}^{\prime}i_{1}^{\prime}}(\lambda_{1}i_{1})\\
  &+\frac{1}{2}\sum\limits_{i^{\prime}}U_{\lambda_{2}i_{2}}^{\lambda_{1}i_{1}}(Ji^{\prime})
  R_{i^{\prime}}(J\nu)=0,
  \label{3pheq2}
  \end{split}
\end{equation}
\begin{equation}
 T_{J^{\prime}{\,}\lambda_{3}^{\prime}i_{3}^{\prime}}^{\lambda_{1}^{\prime}i_{1}^{\prime}{\,}
 \lambda_{2}^{\prime}i_{2}^{\prime}}(J\nu)
 \left(\omega_{\lambda_{1}^{\prime}i_{1}^{\prime}}+\omega_{\lambda_{2}^{\prime}i_{2}^{\prime}}
 +\omega_{\lambda_{3}^{\prime}i_{3}^{\prime}}-E_{\nu}\right)
 +\sum\limits_{i}P_{\lambda_{3}^{\prime}i_{3}^{\prime}}^{J^{\prime}i}(J\nu)
 U_{\lambda_{2}^{\prime}i_{2}^{\prime}}^{\lambda_{1}^{\prime}i_{1}^{\prime}}(J^{\prime}i)=0.
\label{3pheq3}
\end{equation}
The rank of the set of linear Eqs.~(\ref{3pheq1}),
(\ref{3pheq2}) and (\ref{3pheq3}) is equal to the number of one-,
two-, and three-phonon configurations included in the wave
function~(\ref{wf3ph}). To resolve this set it is required
to compute the coupling matrix elements
\begin{equation}
 U_{\lambda_{2}i_{2}}^{\lambda_{1}i_{1}}(Ji)=
 \langle0|Q_{Ji}\mathcal{H}\left[Q_{\lambda_{1}i_{1}}^{+}
 Q_{\lambda_{2}i_{2}}^{+}\right]_{J}|0\rangle
\end{equation}
between one- and two-phonon configurations (see details in~\cite{Severyukhin2004}).
Evidently, the nonzero matrix elements
$U_{\lambda_{2}i_{2}}^{\lambda_{1}i_{1}}(Ji)$ result in the inclusion
of the PPC effects. Equations~(\ref{3pheq1}), (\ref{3pheq2}) and (\ref{3pheq3})
have the same form as the QPM equations~\cite{Grinberg1998}.
However, the SP spectrum and the parameters of
the residual interaction are calculated with
the chosen Skyrme forces, without any further adjustments.
We consider widely used SLy5 EDF~\cite{Chabanat1998} which is adjusted to
reproduce the nuclear matter properties, as well as nuclear charge radii,
binding energies of doubly-magic nuclei~\cite{Bender2003}.

The excitation operator of the ISGMR is defined as
\begin{equation}
 \hat{M}_{\lambda=0}=\sum\limits_{i=1}^{A}r^2_i.
\end{equation}
The wave functions~(\ref{wf3ph}) allow us to determine the transition
probabilities
$\left|\langle0^{+}_{\nu}|\hat{M}_{\lambda=0}|0^{+}_{\text{g.s.}}\rangle\right|^2$.
To construct the wave functions~(\ref{wf3ph}) of the excited $0^{+}$~states,
in the actual calculations~(PPC3), we take into account all two- and
three-phonon configurations below 30~MeV that are built from the phonons
with different multipoles $\lambda^{\pi}=0^{+},1^{-},2^{+},3^{-},4^{+}$,
and $5^{-}$ coupled to~$J^{\pi}=0^{+}$. If we omit the three-phonon configurations,
then this calculation is hereafter called PPC2.

%
\begin{table}[t!]
\caption[]{ Energies and $B(E\lambda)$ values for the transitions
to the ground states in $^{48}$Ca; experimental data
are taken from~\cite{Chen2022}}
\label{tab1}
\begin{tabular}{c|c|c|c|c}
\hline
 $\lambda^{\pi}_{1}$&\multicolumn{2}{c|}{Energy,}&\multicolumn{2}{c}{$B(E\lambda;\lambda^{\pi}_{1}\rightarrow0^{+}_{\text{g.s.}}),$}\\
          &\multicolumn{2}{c|}{MeV} &\multicolumn{2}{c}{W.u.} \\\cline{2-5}
          & Expt.  &  Theory & Expt.        &  Theory         \\
\hline
 $2^{+}_{1}$&3.832 & 3.19    & $1.84_{-0.14}^{+0.17}$& 1.3 \\
 $3^{-}_{1}$&4.507 & 4.47    & $8.4_{-3.5}^{+4.3}$   & 4.1 \\
 $4^{+}_{1}$&4.503 & 3.51    &                       & 2.1 \\
 $5^{-}_{1}$&5.729 & 4.52    &                       & 8.7 \\
\hline
\end{tabular}
\end{table}
Another important aspect in the RPA basic is related to the description
of the low-lying states. The energies and reduced transition probabilities of
the $[2_{1}^{+}]_{\text{RPA}}$, $[3_{1}^{-}]_{\text{RPA}}$, $[4_{1}^{+}]_{\text{RPA}}$,
and $[5_{1}^{-}]_{\text{RPA}}$ states are presented in Table~\ref{tab1}.
The RPA results obtained within the SLy5 EDF are compared with the
experimental data~\cite{Chen2022}. As one can see, the overall
agreement of the energies and $B(E\lambda)$ values with the data
looks reasonable. It is remarkable that the PPC3 plays a minor role, for example,
in the $2_{1}^{+}$~state's description~\cite{Severyukhin2021,Arsenyev2017}.
The crucial contribution to the wave function comes from the neutron configuration
$\{1f_{\frac{7}{2}},2p_{\frac{3}{2}}\}$.

\section{The results of microscopic calculations}

\begin{figure*}[t!]
\includegraphics[width=15cm]{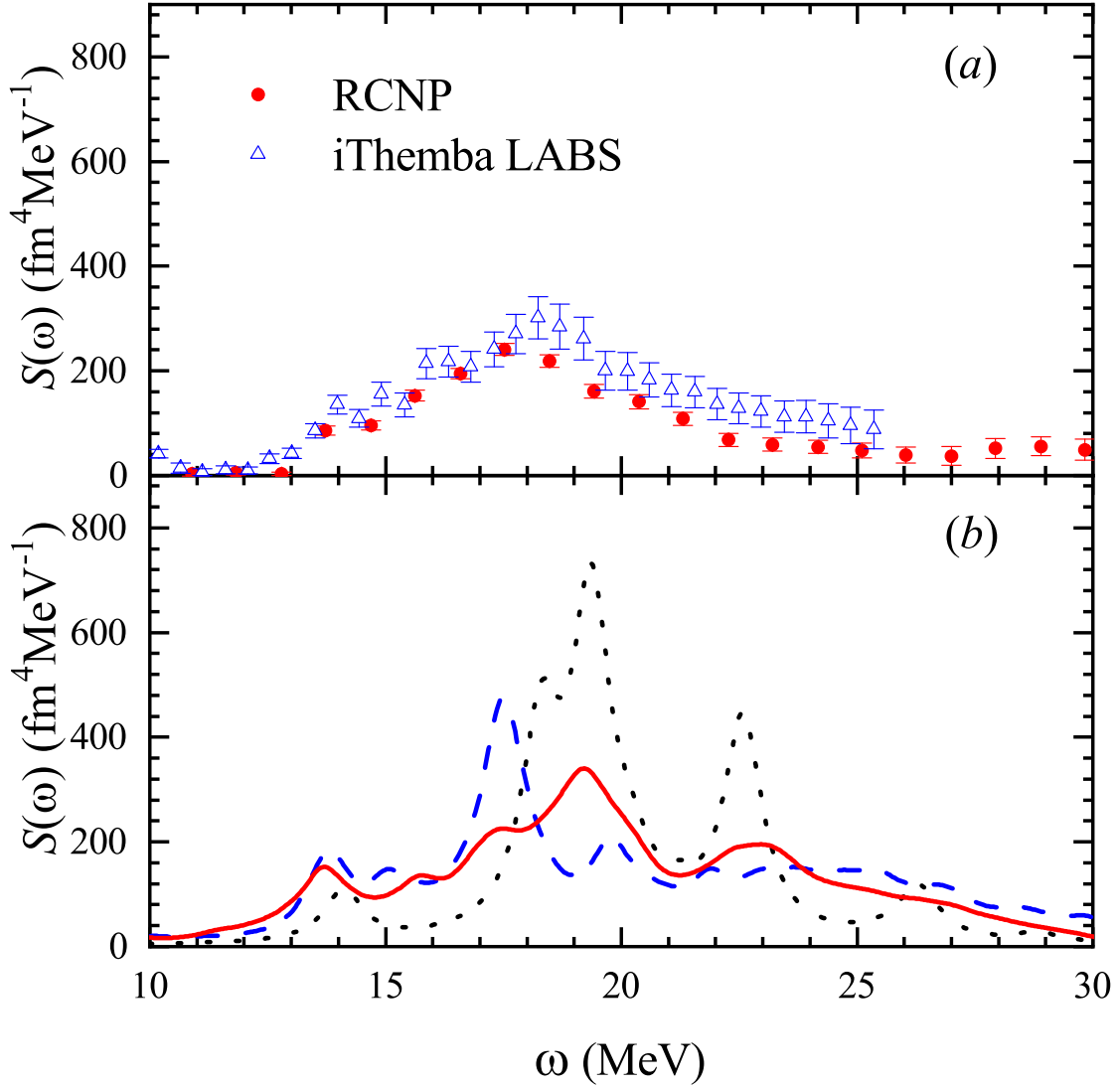}
\caption{Panel~({\it a}) experimental ISGMR strength
distributions. Experimental data are taken
from~\cite{Howard2020,Olorunfunmi2022}. Panel~({\it b}).
Comparison of the results obtained by means of the
microscopic calculations. The dotted, dashed and solid curves
correspond to the RPA, PPC2 and PPC3 calculations, respectively.
The smoothing parameter 1~MeV is used for the strength distribution
described by the Lorentzian function.}
\label{fig1}
\end{figure*}

As a first step in testing our approach we consider the
$E0$~strength distribution in the ISGMR energy region.
Using the excitation energies $E_{\nu}$ and the transition probabilities
$\left|\langle0^{+}_{\nu}|\hat{M}_{\lambda=0}|0^{+}_{\text{g.s.}}\rangle\right|^2$,
the isoscalar monopole strength distribution is averaged out by
a Lorentzian distribution with a width of 1.0~MeV as follows:
\begin{equation}
 S(\omega)=\sum\limits_{\nu}\left|\langle0^{+}_{\nu}|\hat{M}_{\lambda=0}|0^{+}_{\text{g.s.}}\rangle\right|^2
 \frac{1}{2\pi}\frac{\Delta}{(\omega-E_{\nu})^2+\Delta^{2}/4}.
\end{equation}
The ISGMR strength distribution up to 30~MeV is displayed in Fig.~\ref{fig1}.
Unlike the ISGMR for the heavy nuclei ($A>90$)~\cite{Garg2018} which concentrates
in a single collective peak, the strength function of ISGMR for $^{48}$Ca
is fragmented. Both experimental~\cite{Howard2020,Olorunfunmi2022}
and theoretical results showed the fragmentation and splitting of the ISGMR strength.
Let us look at the calculated strength distributions in more detail.
There is the low-energy part below 10~MeV, the main peak in
the ISGMR region (10.0--25.5~MeV) and the high-energy tail above 25.5~MeV.
The results of the RPA calculation are shown in Fig.~\ref{fig1}. According to RPA
calculations the lowest $0^{+}$~state appears above 14~MeV. The
ISGMR strength in the energy range 10.0--25.5~MeV exhausts 92.7{\%}
of the total monopole strength which we obtained below 30~MeV.
The RPA strength distribution is practically concentrated on
three states at 18.3, 19.4, and 22.6~MeV, which exhausts
89.4{\%} of the EWSR, $2\hbar^{2}A\langle{r}^{2}\rangle/m$ \cite{Martorell1976},
or about 79.3{\%} of the total monopole strength. The rest of
the monopole strength lies in the high-energy region.
%

The coupling between the one-, and two-phonon configurations
(dashed curve in Fig.~\ref{fig1}), PPC2, yields a noticeable redistribution
of the ISGMR strength in comparison with the RPA results
(dotted curve in Fig.~\ref{fig1}). Comparing the RPA results, we see that
a single peak appears at 17.4~MeV, which exhausts 13.3{\%} of the EWSR value.
The two-phonon coupling produces a shift of order 0.6{\%} and 8.6{\%}
of the total monopole strength from  the ISGMR region to the low- and
high-energy regions, respectively. We recall that the importance of
the complex configurations for the interpretation of basic peculiarities
of the ISGMR strength distribution of $^{48}$Ca was already qualitatively
discussed in the framework of the microscopic calculations including~$1p1h$
coupled to phonon configurations~\cite{Kamerdzhiev2004}. Our calculations
give the same tendency.

The extension of the configuration space to the three-phonon
configurations (solid curve in Fig.~\ref{fig1}), PPC3, also has
a strong effect on the ISGMR strength distribution in
comparison with the PPC2 results (dashed curve in Fig.~\ref{fig1}).
In particular, the PPC3 effect leads to a shift of the main peak by
1800~keV towards higher energies. Moreover, the three-phonon
coupling induces a shift of order 0.6{\%} of the total monopole strength
from the ISGMR region to the lower energy region ($E_{x}<10.0$~MeV).
At the same time, the PPC3 effect pumps about 5.8{\%} of
the total monopole strength from the high-energy tail of
the ISGMR to the resonance region (10.0--25.5~MeV). As a result,
our analysis showed that the $2p2h$ and $3p3h$~fragmentations shift
about 1.2{\%} (2.8{\%}) of the total monopole strength into
the lower energy region ($E_{x}<10.0$~MeV) and the higher energy region
($E_{x}>25.5$~MeV), respectively. As can be seen from Fig.~\ref{fig1}, the general
shapes of the ISGMR obtained in the PPC3 are somewhat close to those
observed in the RCNP and iThemba LABS experiments. This demonstrates
the improvement of the PPC3 description in comparison with RPA and also PPC2.
We conclude that the main mechanisms of the ISGMR formation in $^{48}$Ca
can be taken into account correctly and consistently in the PPC approach.

%
\begin{table}[t!]
\caption[]{Moment ratios of the $E0$~strength distributions
calculated over the excitation-energy range 10.0--25.5~MeV;
experimental data are taken from~\cite{Olorunfunmi2022}}
\label{tab2}
\begin{tabular}{c|c|c|c}
\hline
          & $m_{1}/m_{0}$, & $\sqrt{m_{1}/m_{-1}}$, & $\sqrt{m_{3}/m_{1}}$, \\
          & MeV            &  MeV                   & MeV\\
\hline
 RPA      & 19.8          & 19.6                    & 20.1 \\
 PPC2     & 18.9          & 18.5                    & 19.8 \\
 PPC3     & 19.1          & 18.8                    & 19.9 \\
 RCNP     & $18.01{\pm}0.10$ & $17.75{\pm}0.10$     & $18.78{\pm}0.13$ \\
 iThemba LABS& $18.40{\pm}0.13$ & $18.09{\pm}0.12$  & $19.29{\pm}0.15$ \\
 TAMU    & $17.52{\pm}0.10$ & $17.11{\pm}0.10$      & $18.69{\pm}0.05$ \\
\hline
\end{tabular}
\end{table}
In order to discuss the various integral characteristics,
we introduce the energy-weighted moments $m_{k}$
\begin{equation}
 m_{k}=\sum\limits_{\nu}\left(E_{\nu}\right)^{k}
 \left|\langle0^{+}_{\nu}|\hat{M}_{\lambda=0}|0^{+}_{\text{g.s.}}\rangle\right|^2.
\end{equation}
These values are useful in estimating the resonance centroid
and also in checking numerical calculations. Let us examine
the various moment rations of the $E0$~strength distribution.
The calculated rations are given in Table~\ref{tab2}. They are
compared with experimental data~\cite{Olorunfunmi2022}.
In~\cite{Olorunfunmi2022}, the experimental datasets of
the three groups are discussed. The integral characteristics of
$E0$~strength function have been obtained in the energy region of
10.0--25.5~MeV. It can be seen, that the PPC3 induces a 700-keV downward
shift of the ISGMR centroid energy $E_{c}=m_{1}/m_{0}$ compared to the RPA;
see Table~\ref{tab2}. In general, these results indicate that the PPC3 predictions
lead to a better agreement (compared to RPA) with these known
experimental values. To test futher the validity of our microscopic approach
we compare the calculated spreading width
$\Gamma=2.35\sqrt{m_{2}/{m_{0}}-\left(m_{1}/{m_{0}}\right)^{2}}$
with the experimental data available for $^{48}$Ca. We obtained that
the PPC3 coupling increases the ISGMR width from 5.1 to 7.8~MeV.
The experimental ISGMR width is $6.68_{-0.36}^{+0.31}$~MeV~\cite{Lui2011},
but we note that in the presented rms width was obtained in
the energy region 9.5--40.0~MeV. In~\cite{Olorunfunmi2020a},
the ISGMR width for $^{48}$Ca is 7.78~MeV obtained through small
angle inelasting $\alpha$-scattering measurements at 200~MeV.
The ISGMR width was extracted by applying a Lorentzian curve in the excitation
energy region of 12.0--24.0~MeV. One can see more and less good agreement
between the calculated and available experimental data.

%
%
\section{Conclusions}

Beginning with the mean-field calculations with the Skyrme
force SLy5 the properties of the spectrum of $0^{+}$~excitations
of $^{48}$Ca were studied within the FRSA model, including the effects
of the phonon-phonon coupling. The inclusion of the coupling between one-, 
two-, and three-phonon terms in the wave functions of excited $0^{+}$~states 
leads to a redistribution of the strength of the ISGMR. A part of the main 
peak strength is fragmented in the low-energy states and the main peak 
itself is shifted downwards. The coupling with $3p3h$~excitations shifts 
about 1{\%} of the total monopole strength into the lower energy region
($E_{x}<10.0$~MeV) and 3{\%} into the higher energy region ($E_{x}>25.5$~MeV).
The crucial contribution in the wave-function structure
of the low-lying $0^{+}$~states ($E_{x}<10.0$~MeV) comes from
the two-phonon configurations. It is shown that the $2p2h$ and $3p3h$~fragmentations
lead to an increase in the ISGMR width. Our results are in reasonable agreement
with the recently experimental data for $^{48}$Ca.

%
%
\section*{Acknowledgments}

Very fruitful discussions with Sunday Olorunfunmi, and Armand Bahini
are gratefully acknowledged. We thank Iyabo Usman for providing insights
and numerical data for the experimental monopole strength in $^{48}$Ca.
A.N.N. acknowledges support from the Russian Science Foundation (Grant No. RSF-21-12-00061).
The work of A.P.S. was supported by the National Research Foundation of
South Africa (Grant No.~129603).

%
%

%

\end{document}